\documentclass[twocolumn,showpacs,preprintnumbers,amsmath,amssymb,aps,prb]{revtex4}
\usepackage{graphicx}
\begin{document}

\title{
  Magnus-Induced Dynamics of Driven Skyrmions on a Quasi-One-Dimensional
  Periodic Substrate  
 } 
\author{
  C. Reichhardt
  and  C. J. Olson Reichhardt 
} 
\affiliation{
Theoretical Division,
Los Alamos National Laboratory, Los Alamos, New Mexico 87545 USA\\ 
} 

\date{\today}
\begin{abstract}

We numerically examine driven skyrmions
interacting with a periodic quasi-one dimensional substrate where the driving 
force is applied either parallel or perpendicular
to the substrate periodicity direction.
For perpendicular driving,
the particles in a purely overdamped system
simply slide along the substrate minima;
however, for skyrmions where the Magnus force is relevant, 
we find that a rich variety of dynamics can arise. 
In the single skyrmion limit,
the skyrmion motion is locked along the driving
or longitudinal direction for low drives,
while
at higher drives a transition occurs to a state in which the
skyrmion moves both transverse and longitudinal to the driving direction.
Within the longitudinally locked phase
we find a
pronounced speed up effect
that occurs
when the Magnus force aligns with the external driving force, while
at the transition to transverse and longitudinal motion,
the skyrmion velocity drops,
producing negative differential conductivity.
For collectively interacting skyrmion assemblies, the speed up effect 
is still present and we observe a number of distinct
dynamical phases, including
a sliding smectic phase, a disordered or moving liquid phase,
a moving hexatic phase, and a moving crystal phase.  The transitions
between the dynamic phases produce distinct features
in the structure of the skyrmion lattice and in the velocity-force curves. 
We map these different phases as a function of the
ratio of the Magnus term to the dissipative term,
the substrate strength, the commensurability ratio,
and the magnitude of the driving force.
\end{abstract}
\maketitle

\section{Introduction}

There are numerous examples of systems that can be
described as individual particles
or a collection of particles interacting with a periodic quasi-one dimensional (q1D)
substrate, including colloids
on optically created q1D substrates \cite{1,2,3,4} or q1D line pinning \cite{5,6,7},
vortices in type-II superconductors with one-dimensional (1D) periodic thickness modulations
\cite{8,9,10,11,12,13,14}, and various frictional systems \cite{15}.
In the colloidal systems a variety of
commensurate-incommensurate states can occur
such as crystal, smectic, and disordered structures \cite{1,2,3,4,5}.
In vortex systems,
under an applied driving force
a series of peaks or dips in the critical depinning force can appear
which are also associated with commensuration effects \cite{8,9,10,12}.
These systems, as well as other systems of particles interacting with
two-dimensional (2D) periodic
substrates,
can exhibit a variety of depinning phenomena and dynamic phases,
and can undergo transitions
between different types of dynamic phases
that produce
changes in the configuration of particles and flow behavior as well as features in the
velocity-force curves
\cite{7,8,14,15,16,17,18,19,20,21,22,23}.
In a 2D overdamped system with a q1D periodic substrate,
pinning-depinning phenomena
and distinct dynamical phases appear only
when the driving force is applied parallel to the substrate periodicity direction.
If the drive is applied perpendicular to the substrate periodicity direction,
there is no pinning effect from the substrate and the particles
simply slide along the driving direction, 
resulting in a linear velocity-force curve.  

In overdamped systems in the absence of a substrate,
individual particles move
in the same direction as the applied driving force.
In some systems, additional transverse forces can arise when a
Magnus force term
${\bf F}_{M}$ with the form 
${\hat z}\times {\bf v}$ is present, which causes
a rotation of the particle velocity into the direction perpendicular to the net
applied forces.
When particles with a Magnus force term are driven
perpendicular to  the periodicity direction of a
q1D periodic substrate, 
the Magnus term creates a coupling between the motion of the particles
parallel and perpendicular to the driving force,
so the effect of the q1D pinning becomes relevant.
A Magnus term can arise for vortices in superconductors and
superfluids; however, in the case of superconducting systems
it is normally very small and has little effect
on the depinning and sliding dynamics.
Recently a new particle-like system, skyrmions in chiral magnets, was discovered
in which the Magnus force is much stronger
\cite{34,35,36,37}.
Since the initial observation of skyrmions in magnetic systems, there has been a rapid
growth in the field as an increasing number of systems have been
identified that support skyrmions,
including materials 
in which skyrmions are stable at room temperature \cite{38,39,40,41,42}. 
Another reason for the growing interest in this field is that skyrmions could be
used as magnetic information carriers, making them promising
for spintronic applications \cite{43}.

In order for spintronic or other applications
of skyrmions to be realized, it is necessary to have an understanding of how skyrmions
move in different types of  nanostructured samples.
Skyrmions can be moved by an applied current \cite{37,M,J},
and have been shown to exhibit a pinned to sliding transition based on
effective velocity-force curves that can be constructed by measuring
changes in the transport properties \cite{44,45}.
Velocity-force curves can also be obtained by directly imaging the skyrmion motion
\cite{37,38,46}.
In many cases the skyrmion critical depinning force
is very
low, and this was
argued to be a result of the Magnus term which permits skyrmions
to move around a pinning site and avoid trapping rather than
moving toward the pinning site
and becoming trapped as in the case of overdamped systems \cite{47,48,49}.
Under an applied driving force, the Magnus term causes the skyrmion
to move at an angle with respect to the
driving direction, producing a skyrmion Hall angle $\theta_{sk}$ \cite{37,J,47}.
In a pin-free system
$\theta_{sk}$ is a constant and
is proportional to the ratio of the Magnus term $\alpha_{m}$
to the damping term $\alpha_{d}$,
$\theta_{sk} \propto \tan^{-1}(\alpha_{m}/\alpha_{d})$.
When pinning is present, however,
$\theta_{sk}$
becomes drive-dependent as the skyrmions make
a side jump motion  when interacting with an individual pinning site, which reduces the
Hall angle \cite{49,50,51}.
As
the drive is increased, the side jump effect is reduced and
$\theta_{sk}$
approaches the clean value limit.
In Ref.~\cite{52} an imaging technique provided
direct evidence for the drive dependence of the skyrmion Hall angle,
with a linear
dependence of the ratio of the
transverse to longitudinal skyrmion velocity as a function of drive.
These studies have focused on point-like pinning or circular pinning sites;
however, it should also be possible
to create line-like pinning using various lithographic techniques
such as 1D periodic thickness modulation, periodic magnetic
strips, or
optical techniques.

In this work we use a particle-based simulation to examine individual and
collectively interacting skyrmions in a 2D system in the
presence of a q1D periodic substrate, as
described Section II.
The particle model is based
on a modified Thiele equation \cite{48,49,50,51,J1}
which agrees well with continuum-based simulations in the limit where
the overlap of adjacent skyrmions
is small \cite{48}.
In Section III we describe the results for the single skyrmion limit,
where if the drive is applied parallel to the substrate
periodicity direction, we find that unlike the case of point-like pinning,
the depinning threshold does not decrease
with increasing Magnus term magnitude since the skyrmions cannot simply
move around the pinning sites.
When the drive is applied perpendicular to the substrate
periodicity direction, in the overdamped limit there is no depinning
threshold and
the skyrmions simply slide without any structural change for
increasing drive, producing a linear velocity-force curve.
On the other hand, when there is
a finite Magnus term we observe a
rich variety of dynamical behaviors even in the single skyrmion limit.
For perpendicular drives,
the skyrmion motion is locked in the drive direction
at low drives until a critical driving force is reached at which
the skyrmions also start move partially parallel to the substrate
periodicity direction,
coinciding with a sudden drop in the net velocity of
the skyrmion and producing a negative differential
conductivity effect. At higher drives the skyrmion velocity again increases
with increasing drive.
We also show that when the skyrmion motion
is locked in the direction of the drive, a speed up effect occurs where
the skyrmion moves faster than it would in the overdamped limit
due to the alignment of the
pinning-induced velocity from the Magnus term with the driving force direction.
This speed up effect
was initially observed in simulations of point-like pinning \cite{49,50};
however, the effect is more easily controlled with q1D
pinning.
When the driving is applied
parallel to the substrate periodicity direction, 
there is no speed up effect but instead an enhanced damping appears.
In Section IV we examine
collectively interacting skyrmions and show that the same speed up effect
and transition from locked motion in the direction of drive to motion in both the
longitudinal and transverse directions occur.  In addition,
a series of 
dynamical phases appear that can be characterized by the structure of the
moving skyrmions, 
and the transitions between these phases
are correlated with distinct features in the transport curves. 
The phases include a moving smectic and a moving liquid
which can undergo dynamical ordering transitions into
a moving quasi-ordered hexatic lattice or a moving crystal.
We map out the dynamic phases as a function of the substrate strength and the
ratio of
the Magnus force to the dissipative term.
In Section V we examine the effect of changing the ratio
of the skyrmion density to the periodicity of the  substrate,
where we observe chainlike structures consisting of
multiple rows of skyrmions per substrate minimum.
We also check for hysteresis across the dynamic phase transitions.

\begin{figure}
\includegraphics[width=3.5in]{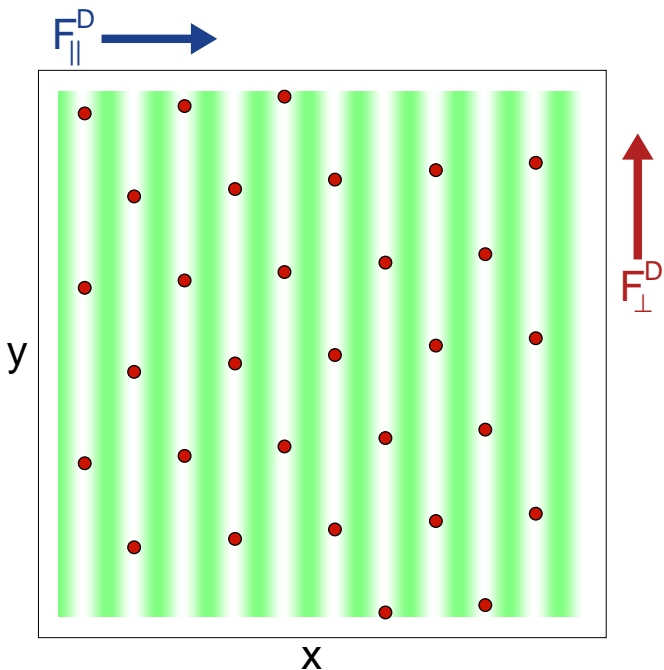}
\caption{
(Color online) Skyrmions (red dots) at a density of
  $\rho_s=0.1$ on a periodic quasi-one-dimensional substrate
  with pinning strength $A_p=1.0$. Here the
  substrate periodicity is in the $x$-direction
  and we consider dc driving $F^D_{||}$ in the parallel or $x$ direction (blue arrow)
  and $F^D_{\perp}$ in the perpendicular or $y$ direction (red arrow).
  The dark green regions indicate the locations of the substrate potential
  maxima.
}
\label{fig:1}
\end{figure}

\section{Simulation}   
In Fig.~\ref{fig:1} we show a snapshot of our 2D system, which has
periodic boundary conditions in the $x$ and $y$ directions and contains a
q1D periodic sinusoidal substrate potential with a period $a$
and a periodicity running along the $x$ direction.
There are $N$ skyrmions which are trapped in the potential minima.
The initial skyrmion positions are obtained through simulated annealing,
after which we apply a dc driving force ${\bf F}^{dc}$ in either
the parallel or $x$ direction, ${\bf F}^{dc}=F^D_{||}{\bf \hat x}$,
or in the perpendicular or $y$ direction, ${\bf F}^{dc}=F^D_{\perp}{\bf \hat y}$,
and we measure the  resulting
skyrmion velocity.   
The dynamics of a single skyrmion $i$ are obtained using the following
equation of motion:
\begin{equation}
  \alpha_{d}{\bf v}_{i} + \alpha_{m}{\hat z}\times {\bf v}_{i} =
        {\bf F}^{ss}_{i} +  {\bf F}^{sp}_{i}  + {\bf F}^{dc} 
\end{equation}
where ${\bf r}_i$ is the skyrmion position and ${\bf v}_i=d{\bf r}_i/dt$ is the
skyrmion velocity.
The first term is the damping
$\alpha_{d}$
which  aligns the skyrmion velocity in the direction
of the net external forces, and the second term is the
Magnus force with prefactor $\alpha_{m}$,
where the cross product creates a velocity component perpendicular to the
net external forces.
To  maintain a constant magnitude of the skyrmion velocity we apply
the constraint $\alpha_{d}^2 + \alpha^2_{m} = 1$.
The skyrmion Hall effect can be characterized
by measuring the ratio
$R = \langle V_{\perp}\rangle/\langle V_{||}\rangle$
of the skyrmion velocity in the perpendicular
direction, $\langle V_{\perp}\rangle = N^{-1}\sum_{i}^{N} {\bf v}_i \cdot {\bf \hat y}$,
to that in the parallel direction, 
$\langle V_{||}\rangle = N^{-1}\sum_{i}^{N} {\bf v}_i \cdot {\bf \hat x}$.
The skyrmion Hall angle is $\theta_{sk} = \tan^{-1}(R)$. 
In a clean system, $R$ has a constant value given by $R=\alpha_{m}/\alpha_{d}$.   
The substrate force
${\bf F}^{sp}_i  = -\nabla U(x_i) {\bf \hat x}$ arises from a periodic sinusoidal potential
\begin{equation}
U(x_i)  = U_{0}\cos(2\pi x_i/a)
\end{equation}
where $x_i={\bf r}_i \cdot {\bf \hat x}$,
$a$ is the periodicity of the substrate, and
we define the substrate strength to be $A_{p} = 2\pi U_{0}/a$.
The skyrmion-skyrmion interaction force is repulsive, which  favors
the formation of a triangular lattice in a clean system.  It has the form
${\bf F}^{ss}_{i} = \sum_{j=1}^{N}K_{1}(R_{ij}){\bf \hat r}_{ij}$
where $R_{ij} = |{\bf r}_{i} - {\bf r}_{j}|$, 
${\hat {\bf r}_{ij}} = ({\bf r}_i - {\bf r}_{j})/R_{ij}$, and
$K_{1}$ is the modified Bessel function. This 
interaction falls off exponentially for large $R_{ij}$.
The sample is of size $L \times L$ and the skyrmion density is $n_s=N/L^{2}$. Previous
studies of skyrmions on a similar q1D periodic substrate focused on Magnus-induced
Shapiro steps, which arise when an additional ac drive is present \cite{1}.

\section{Single Skyrmion Limit}

\begin{figure}
\includegraphics[width=3.5in]{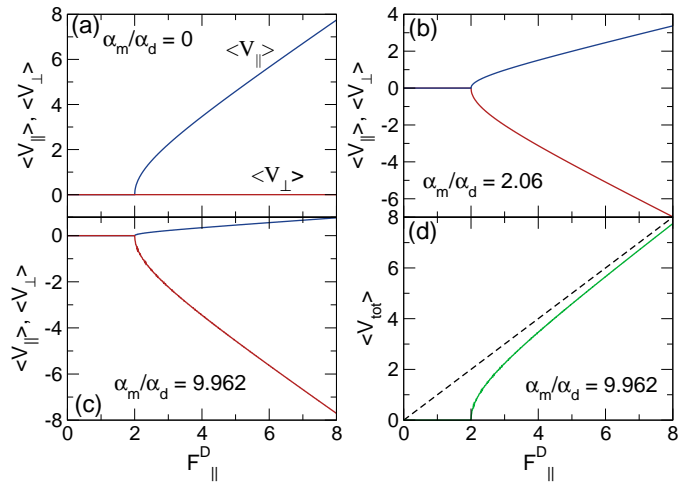}
\caption{ (Color online) Parallel ($\langle V_{||}\rangle$, blue) and perpendicular
  ($\langle V_{\perp}\rangle$, red) velocities for a single skyrmion driven in
  the parallel ($x$) direction vs the driving force magnitude $F^{D}_{||}$.
  The substrate potential strength is $A_p=2.0$.
  (a) In the overdamped limit of $\alpha_{m}/\alpha_{d} = 0$,  the motion
  is locked in the  parallel direction
  and there is a critical depinning force of $F^c_{||} = 2.0$.
  (b) At $\alpha_{m}/\alpha_{d} = 2.06$, there is a finite velocity signal in both directions. (c)
  At $\alpha_{m}/\alpha_{d} = 9.962$, $\langle V_{||}\rangle$ is diminished compared
  to $\langle V_{\perp}\rangle$.
  (d) $\langle V_{\rm tot}\rangle = \sqrt{\langle V_{||}\rangle^2 + \langle V_{\perp}\rangle^2}$
  vs $F^D_{||}$ for $\alpha_m/\alpha_d=9.962.$  The dashed line indicates the
response $\langle V_{\rm tot}^0\rangle$ for a clean system with $A_p=0.$
The $\langle V_{\rm tot}^0\rangle$ curve does not vary as a function of
$\alpha_m/\alpha_d$.
}   
\label{fig:2}
\end{figure}

\begin{figure}
\includegraphics[width=3.5in]{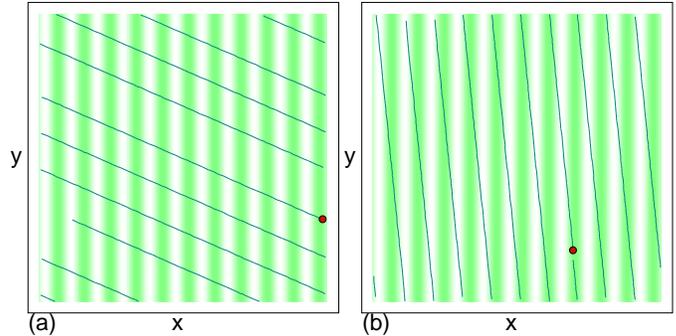}
\caption{The skyrmion location (red circle),
  trajectory (line), and substrate potential (green)
  for the system in Fig.~\ref{fig:2} just above depinning.
  The dc drive $F^D_{||}$ is in the positive $x$-direction.
  (a) At $\alpha_{m}/\alpha_{d} = 0.4364$, the skyrmion moves at an
  angle of $\theta_{sk} = 23.6^{\circ}$ with respect to the driving direction.
(b) At $\alpha_{m}/\alpha_{d} = 9.9624$, $\theta_{sk} = 84.26^{\circ}$.}  
\label{fig:3}
\end{figure}

We first consider the case of a single skyrmion, $N=1$.
In Fig.~\ref{fig:2}(a) we plot $\langle V_{||}\rangle$
and $\langle V_{\perp}\rangle$ versus
$F^{D}_{||}$
for an overdamped system
with $\alpha_{m}/\alpha_{d} = 0$ where the particle is driven parallel
to the substrate periodicity direction and the substrate strength is
$A_p=2.0$.
Here, $\langle V_{\perp}\rangle = 0$ for all $F^{D}_{||}$, and
there is a depinning transition at $F^{c}_{||} = A_{p} = 2.0$,
above which $\langle V_{||}\rangle$ becomes finite.
Figure~\ref{fig:2}(b) shows that at
$\alpha_{m}/\alpha_{d} = 2.06$,
the depinning threshold is still $F^{c}_{||} = 2.0$, but the skyrmion
now moves both parallel and perpendicular to the driving direction
above depinning.
The slope of the $\langle V_{\perp}\rangle$ curve is approximately twice that of
the $\langle V_{||}\rangle$ curve.
At $\alpha_{m}/\alpha_{d} = 9.962$, 
Fig.~\ref{fig:2}(c) shows that the depinning threshold is unchanged
at $F^{c}_{||} = 2.0$
but that the perpendicular velocity has become much more pronounced.
We find that $F^{c}_{||}$ is independent
of $\alpha_{m}/\alpha_{d}$ for driving in the parallel direction.
This is in contrast to observations of skyrmions
interacting with randomly placed \cite{47,48,51} or periodic \cite{49} arrays
of pointlike pinning sites, where
$F^{c}_{||}$ decreases with increasing $\alpha_{m}/\alpha_{d}$.
For pointlike pinning, as the increasing Magnus term causes the
skyrmion trajectories to become increasingly curved,
the skyrmions can more easily
circle around the pinning sites without becoming trapped,
and this has been argued to be one of the reasons
that the depinning thresholds are so low in skyrmion systems.
In the case of the q1D periodic substrate, the pinning potential is planar in one direction,
making it impossible for the skyrmions to circle around the pinning locations.
As a result, planar or linelike pinning sites produce much stronger skyrmion pinning
than pointlike pinning sites.

In Fig.~\ref{fig:2}(d) we plot the total velocity
$\langle V_{\rm tot}\rangle = \sqrt{\langle V_{||}\rangle^2 + \langle V_{\perp}\rangle^2}$
for the system in Fig.~\ref{fig:2}(c) with $\alpha_m/\alpha_d=9.962$.
The dashed line indicates the response $\langle V_{\rm tot}^0\rangle$ in a clean system
with $A_p=0$ for comparison.
Here, for any finite value of $A_{p}$,
$\langle V_{\rm tot}\rangle <  \langle V_{\rm tot}^0\rangle$ for
parallel driving,
indicating that the effective damping is enhanced by the substrate.
In Fig.~\ref{fig:3}(a) we plot the skyrmion trajectory just above depinning
for the system in Fig.~\ref{fig:2} at
$\alpha_{m}/\alpha_{d} = 0.4364$.
The skyrmion follows a straight trajectory oriented
at an angle, the skyrmion Hall angle $\theta_{sk}=23.6^\circ$,
with respect to the external drive.
Figure~\ref{fig:3}(b) shows that at $\alpha_{m}/\alpha_{d} = 9.9624$, the skyrmion
moves at a much
steeper angle to the external drive,
with $\theta_{sk}$ just under the clean limit value
of $\theta_{sk}=84.26^{\circ}$.

\begin{figure}
\includegraphics[width=3.5in]{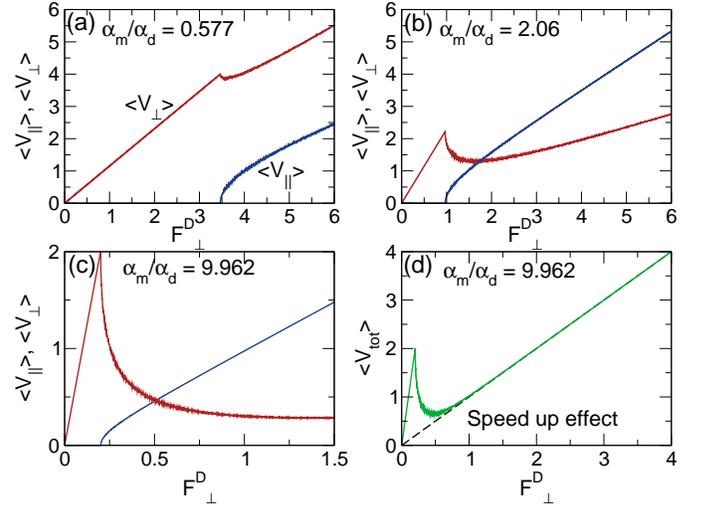}
\caption{
  $\langle V_{||}\rangle$ and $\langle V_{\perp}\rangle$ for a single
  skyrmion driven in the perpendicular ($y$) direction vs the driving force
  magnitude $F^D_{\perp}$.
  The substrate potential strength is $A_{p} = 2.0.$
  (a) At $\alpha_{m}/\alpha_{d} = 0.577$,
  the skyrmion motion is initially locked in the perpendicular direction,
  and a transition to motion in both the parallel and perpendicular directions occurs at
$F^{c}_{\perp} = 3.5$. (b) At $\alpha_{m}/\alpha_{d} = 2.06$,
  $F^{c}_{\perp}$ is decreased. (c)
  At $\alpha_{m}/\alpha_{d} = 9.962$, $F^c_{\perp}$ is even smaller.
  (d) $\langle V_{\rm tot}\rangle $ vs $F^{D}_{\perp}$, where the dashed
  line indicates the response $\langle V_{\rm tot}^0\rangle$ in a system with $A_{p} = 0$.
  Here $\langle V_{\rm tot}\rangle >\langle V_{\rm tot}^0\rangle$, indicating
  the existence of a speed up effect.         
}
\label{fig:4}
\end{figure}

\begin{figure}
\includegraphics[width=3.5in]{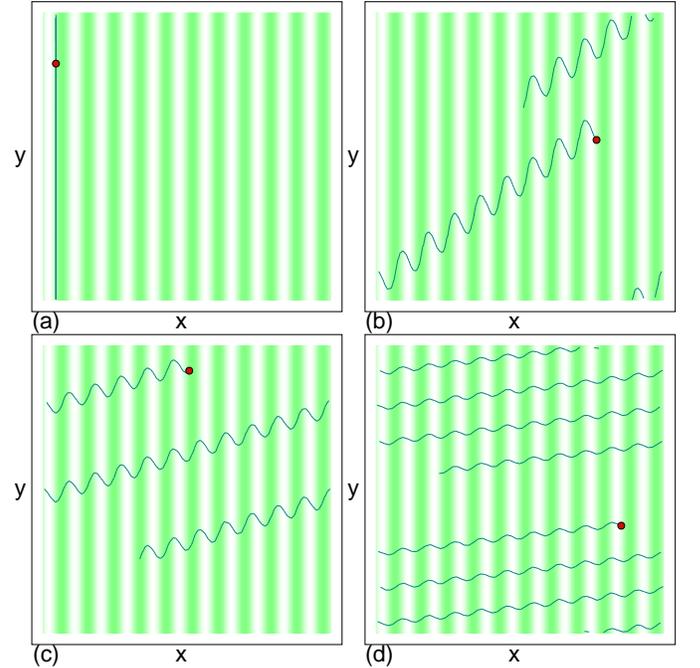}
\caption{ The skyrmion location (red circle), trajectory (line),
  and substrate potential (green) for the system in
  Fig.~\ref{fig:4}(c,d) at $\alpha_{m}/\alpha_{d} = 9.962$.
  The dc drive $F^D_{\perp}$ is in the positive $y$-direction.
(a) At $F^{D}_{\perp} = 0.17$,  the motion is locked in the driving direction. 
  (b) At $F^{D}_{\perp} = 0.6$,  the skyrmion is moving at an angle with respect
  to the dc drive. (c) $F^{D}_{\perp} = 1.0$.
(d) $F^{D}_{\perp} = 3.0$.  
}
\label{fig:5}
\end{figure}

In Fig.~\ref{fig:4} we show $\langle V_{\perp}\rangle$ and
$\langle V_{||}\rangle$ versus $F^D_{\perp}$ for a single skyrmion driven
along the $y$ direction, perpendicular to the substrate symmetry direction.
In the overdamped limit of $\alpha_{m}/\alpha_{d} = 0$,
$\langle V_{||}\rangle = 0$ for all $F^D_{\perp}$, there is no 
depinning threshold for motion in the driving direction, 
and $\langle V_{\perp}\rangle$ increases linearly
with $F^D_{\perp}$.
When $\alpha_{m}/\alpha_{d} > 0$, 
there is a range of $F^{D}_{\perp}$ over which
the skyrmion motion is locked in the perpendicular or $y$ direction, 
and once $F^{D}_{\perp}$ reaches a critical threshold $F^c_{\perp}$,
a transition occurs to motion in both the perpendicular and parallel directions.  This
is illustrated in
Fig.~\ref{fig:4}(a) for $\alpha_{m}/\alpha_{d} = 0.577$,
where $\langle V_{||}\rangle$ becomes finite at
$F^{D}_{\perp} = F^c_{\perp}=3.5$.
This transition 
coincides with a small drop in
$\langle V_{\perp}\rangle$.
As $\alpha_{m}/\alpha_{d}$ is increased,
the value of $F^{c}_{\perp}$ 
decreases while the magnitude of the drop in
$\langle V_{\perp}\rangle$ at the transition  point 
increases, as shown in Fig.~\ref{fig:4}(b,c)
for $\alpha_{m}/\alpha_{d} = 2.06$ and $9.962$, respectively.
For $F^D_{\perp}>F^c_{\perp}$, 
$\langle V_{\perp}\rangle$ increases with increasing $F^{D}_{\perp}$. 
In Fig.~\ref{fig:5}(a) we plot the skyrmion trajectory
for the system in Fig.~\ref{fig:4}(c,d) with $\alpha_{m}/\alpha_{d} = 9.962$
at $F^{D}_{\perp} = 0.17$, where the skyrmion
motion is locked in the drive direction.
At $F^D_{\perp}=0.6$ in Fig.~\ref{fig:5}(b),
the skyrmion moves
in both the longitudinal and transverse directions with a sinusoidal undulation.
In Fig.~\ref{fig:5}(c) at $F^{D}_{\perp} = 1.0$,
the angle $\theta_{sk}$ between the direction of skyrmion motion and the driving direction
is larger, while at $F^D_{\perp}=3.0$ in Fig.~\ref{fig:5}(d),
$\theta_{sk}$ is even larger.  For
high enough drives, $\theta_{sk}$ approaches the clean limit value of
$\theta_{sk}=84.26^{\circ}$.
This shows that the skyrmion Hall angle has a much stronger
dependence on the external driving force for perpendicular driving than for parallel
driving.

In Fig.~\ref{fig:4}(d) we plot the net velocity
$\langle V_{\rm tot}\rangle$ versus $F^D_{\perp}$
at $\alpha_{m}/\alpha_{d} = 9.962$. 
The dashed line shows the response $\langle V_{\rm tot}^0\rangle$ for a system with
$A_{p} = 0$.
Just above the depinning threshold $F^c_{\perp}=0.2$ for motion in the
parallel direction,
we observe
negative differential conductivity (NDC),
where the net velocity of the skyrmion decreases with increasing drive.
Negative differential conductivity is a phenomenon often found for charge 
transport in semiconductors \cite{53},   
and it can be useful in constructing logic devices,
which suggests that the construction of magnetic versions of semiconductor
logic devices using skyrmions may be possible. 
NDC has also been observed for vortices in type-II superconductors driven over periodic
pinning arrays, where it is associated with transitions in the flow states \cite{18,20}.
Previous simulation studies of skyrmions driven over 
2D periodic arrays, where the skyrmion Hall angle changes as a function of drive,
also showed NDC \cite{49}, while in both particle-based and continuum simulations of
skyrmions interacting with an isolated circular pinning site,  the skyrmion
velocity can drop to zero at high enough drive when it becomes possible for
the pinning site to capture a skyrmion \cite{50}.

\begin{figure}
\includegraphics[width=3.5in]{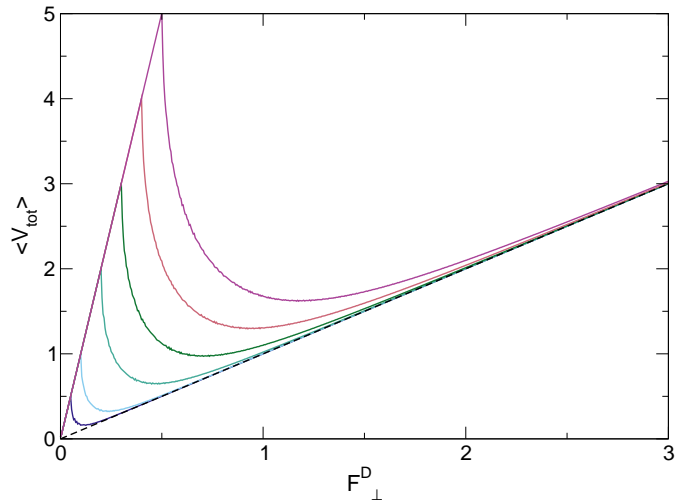}
\caption{ $\langle V_{\rm tot}\rangle$ vs
  $F^{D}_{\perp}$ at
    $\alpha_{m}/\alpha_{d} = 9.962$  for
  $A_{p} = 0.5$, 1.0, 2.0, 3.0, 4.0, and $5.0$, from bottom to top.
The dashed line is $\langle V_{\rm tot}^0\rangle$ with $A_{p} = 0.$
}
\label{fig:6}
\end{figure}

Figure~\ref{fig:4}(d) also shows that
$\langle V_{\rm tot}\rangle$ is always larger than the
clean limit value of $\langle V_{\rm tot}^0\rangle$,
indicating that the q1D substrate enhances the skyrmion
velocity compared to the clean limit.
For example, at the parallel depinning transition point $F^D_{\perp}=F^c_{\perp}=0.2$,
$\langle V_{\rm tot}\rangle \approx 2.0$,
while in the
pin-free limit, $\langle V_{\rm tot}^0\rangle=F^D_{\perp}=0.2$.
Such speed up effects were
first observed in continuum  and particle based
simulations for skyrmions interacting with a single pinning site \cite{50} and
with a periodic array of pinning sites \cite{49}.
In the case of q1D planar pinning sites, it is easier to see that this
effect arises due to the Magnus force.
Due to the damping term, the perpendicular external drive $F^D_{\perp}$ 
produces a perpendicular velocity component of
$\langle V^d_{\perp}\rangle=\alpha_dF^D_{\perp}$.
The Magnus term transfers some of the motion produced by the drive into
the parallel direction, giving a finite value of $\langle V_{||}\rangle$;
however, the pinning imparts
a force proportional to $A_p$ on the skyrmion in the parallel direction.
This parallel force is transformed by the Magnus term into a perpendicular
velocity, leading to an additional velocity contribution
of $\langle V^m_{\perp}\rangle=\alpha_mA_p$.
As long as the skyrmion motion remains locked in the perpendicular direction,
its maximum perpendicular velocity is given by
\begin{equation}
\langle V^{\rm max}_{\perp}\rangle =\alpha_{d}F^{c}_{\perp} + \alpha_{m}A_{p}.
\end{equation}
In Fig.~\ref{fig:4}(c,d) we use $\alpha_{d} = 0.09987$ and $\alpha_{m} = 0.995$,
so that at the transition point
$F^{D}_{\perp} = 0.2$ we obtain $\langle V^{\rm max}_{\perp}\rangle = 2.001$,
in agreement with the maximum
values of $\langle V_{\perp}\rangle$ and $\langle V_{\rm tot}\rangle$ in Fig.~\ref{fig:4}(c,d).
This shows that the skyrmion velocity can increase linearly with the pinning force.
Once the system depins in the parallel direction at $F^c_{\perp}$,
the skyrmion experiences an oscillating pinning force,
causing  the speed up effect to be lost and
$\langle V_{\perp}\rangle$ to drop.
At high drives the system gradually approaches the
clean value limit in which the velocity increases
linearly with drive according to
$\langle V_{\perp}\rangle = \alpha_{d}F^{D}_{\perp}$
and $\langle V_{||}\rangle = \alpha_{m}F^{D}_{\perp}$,
as also observed in systems with periodic and random pointlike pinning \cite{51}.
In Fig.~\ref{fig:6} we plot $\langle V_{\rm tot}\rangle$ versus $F^{D}_{\perp}$
for $A_{p} = 0.5$, 1.0, 2.0, 3.0, 4.0, and $5.0$
at
$\alpha_{m}/\alpha_{d} = 9.962$.
Here, both $F^{c}_{\perp}$ and the maximum value of
$\langle V_{\rm tot}\rangle$ increase with $A_{p}$,
in agreement with Eq. 3.

\begin{figure}
\includegraphics[width=3.5in]{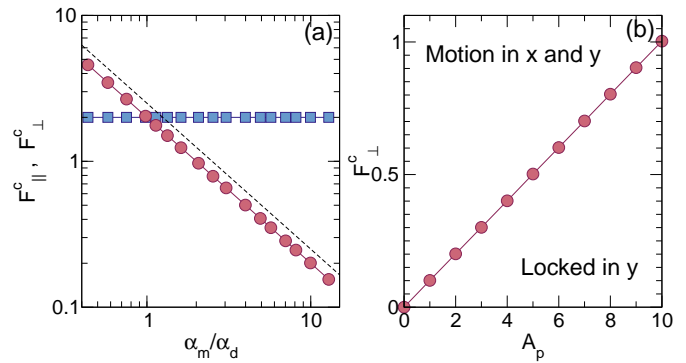}
\caption{(a)
  The depinning forces $F^c_{||}$ and $F^c_{\perp}$ at which motion in the direction
  parallel to the substrate periodicity
  occurs vs $\alpha_m/\alpha_d$ for samples with $A_p=2.0$.
  Blue squares:  $F^{c}_{||}$, for parallel driving,
 has no dependence on $\alpha_m/\alpha_d$.
  Red circles: $F^c_{\perp}$, for perpendicular driving,
  can be fit to
$F^{c}_{\perp} \propto (\alpha_{m}/\alpha_{d})^{-1}$ (dashed line). 
  (b) $F^{c}_{\perp}$ vs $A_{p}$ at
  $\alpha_{m}/\alpha_{d} = 9.962$.
  Here $F^c_{\perp}$ increases linearly with $A_{p}$. 
}
\label{fig:7}
\end{figure}

In Fig.~\ref{fig:7}(a) we plot the depinning force
$F^{c}_{||}$ versus $\alpha_m/\alpha_d$ for parallel driving
for the system in Fig.~\ref{fig:2}
along with the force $F^c_{\perp}$ at which sliding along the parallel direction
occurs for the system in Fig.~\ref{fig:4} with perpendicular driving.
$F^{c}_{||}$ is constant and $F^{c}_{\perp}$ obeys
$F^c_{\perp} \propto (\alpha_{m}/\alpha_{d})^{-1}$,
as indicated by the dashed line, so that $F^{c}_{\perp}$ diverges 
at $\alpha_{m}  = 0$
when the skyrmions stay locked in the direction of drive in the overdamped limit. 
Figure~\ref{fig:7}(b) shows
$F^{c}_{\perp}$ versus $A_{p}$ for
$\alpha_{m}/\alpha_{d} = 9.962$,
showing a linear increase with $A_{p}$ 
with a slope of $A_{p}/\alpha_{m}$.    

\section{Collective Effects}

\begin{figure}
\includegraphics[width=3.5in]{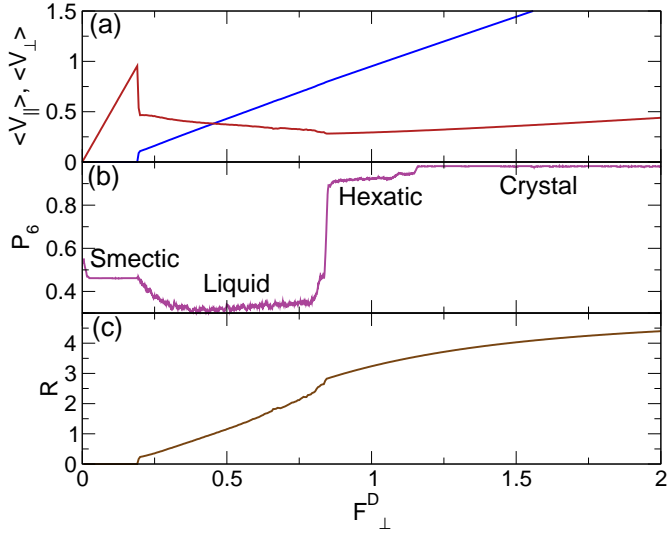}
\caption{
  (a) $\langle V_{||}\rangle$ (blue) and $\langle V_{||}\rangle$
  (red) vs $F^{D}_{\perp}$
  for a system with multiple interacting skyrmions at
  $\alpha_{m}/\alpha_{d} = 4.925$, $A_{p} = 1.0$, 
  and skyrmion density $n_{s} = 0.16$.
  (b) The fraction of sixfold-coordinated skyrmions
  $P_{6}$ vs $F^{D}_{\perp}$.
  In the moving smectic phase the motion is locked along the perpendicular direction.
  Also marked are the moving liquid, moving hexatic, and moving crystal phases.
 (c) $R = \langle V_{\perp}\rangle/\langle V_{||}\rangle$ vs $F^{D}_{\perp}$.
}
\label{fig:8}
\end{figure}

\begin{figure}
\includegraphics[width=3.5in]{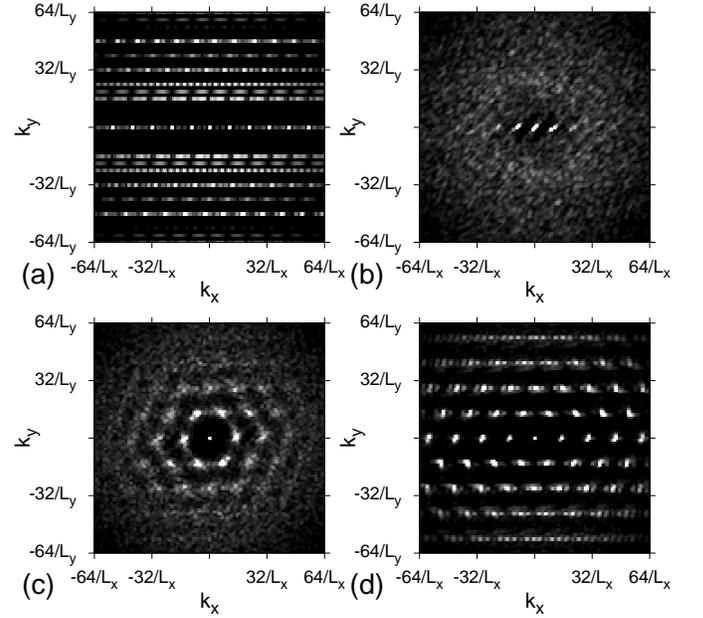}
\caption{The structure factor $S({\bf k})$ for the different phases for the system in
  Fig.~\ref{fig:8}.
  (a) The moving smectic phase at $F^{D}_{\perp} = 0.1$.
  (b) The moving liquid phase at $F^{D}_{\perp} = 0.5$.
  (c) The moving hexatic phase at $F^{D}_{\perp} = 1.0$.
  (d) The moving crystal phase at $F^{D}_{\perp} = 1.2$.
}
\label{fig:9}
\end{figure}

\begin{figure}
\includegraphics[width=3.5in]{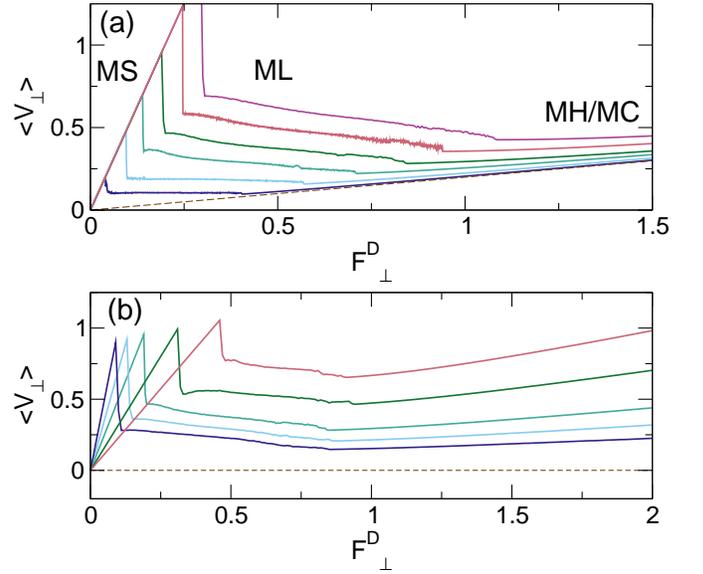}
\caption{(a) $\langle V_{\perp}\rangle$ vs $F^{D}_{\perp}$ for the system in Fig.~\ref{fig:8} 
  at $\alpha_{m}/\alpha_{d} = 4.925$ for $A_{p} = 0.25$, 0.5, 0.75, 1.0,  1.25, and $1.5$,
  from bottom to top,
  showing more clearly the cusp at the ML-MH/MC transition.
  MS: moving smectic; ML: moving liquid; MH: moving hexatic; MC: moving crystal.
  The dashed line is the response for a sample with $A_{p} = 0.$
  (b) $\langle V_{\perp}\rangle$ vs $F^{D}_{\perp}$ for the same system with
  $A_{p} = 1.0$ for 
  $\alpha_{m}/\alpha_{d}   =  9.995$, 7.017, 4.925, 3.0, and $2.06$, from top right
to bottom right.  The dashed line indicates the response in an overdamped system with
 $\alpha_{m}/\alpha_{d} = 0.$
}
\label{fig:10}
\end{figure}

We next consider the case of multiple interacting skyrmions.
In Fig.~\ref{fig:8}(a) we plot $\langle V_{||}\rangle$ and $\langle V_{\perp}\rangle$
versus $F^{D}_{\perp}$
for a system with $\alpha_{m}/\alpha_{d} = 4.925$ and $A_{p} = 1.0$
at a skyrmion density of $n_{s} = 0.16$, where the ratio of the
substrate lattice constant $a$ to the skyrmion lattice constant $a_{sk}$ is
$a/a_{sk} = 1.3098$. Here we observe the same features in
$\langle V_{||}\rangle$ and $\langle V_{\perp}\rangle$
that appeared in the single skyrmion case, including
a longitudinally locked phase,
negative differential conductivity, and a speed up effect.
There are several differences, including additional cusps in the
velocity-force curves at higher drives
which are correlated with changes in the collective dynamics.
Fig.~\ref{fig:8}(b)
shows $P_{6}$, the fraction of sixfold-coordinated skyrmions, versus $F^{D}_{\perp}$.
Here $P_6=N^{-1}\sum_{i=1}^{N}\delta(z_i-6)$, where $z_i$ is the coordination number
of skyrmion $i$ obtained from a Voronoi construction.
In the longitudinally locked regime, the pinning is strong enough that
the skyrmions form 1D incommensurate chains
moving in the perpendicular direction,
so that the skyrmion lattice structure exhibits a number of aligned
topological defects.
Figure~\ref{fig:9}(a) shows the structure factor
$S({\bf k})=N^{-1}|\sum_i^N \exp(-i{\bf k} \cdot {\bf r}_i)|^2$ for the moving
smectic (MS) phase at $F^{D}_{\perp}  = 0.1$,
where the system forms stripe like features indicative of smectic ordering.
For
$0.18 < F^{D}_{\perp} < 0.84$, $\langle V_{\perp}\rangle$  gradually decreases
with increasing $F^D_{\perp}$ and the skyrmions form a disordered or
moving liquid (ML) state, as indicated by the
ring structure in Fig.~\ref{fig:9}(b), which shows
$S({\bf k})$
at $F^{D}_{\perp} = 0.5$.
There are still two satellite peaks in $S({\bf k})$ along the $k_{y} = 0$ line
that are produced by the 1D periodicity of the substrate.

In Fig.~\ref{fig:8}, a cusp appears in $\langle V_{\perp}\rangle$
near $F^{D}_{\perp} = 0.85$, above which $\langle V_{\perp}\rangle$ starts
to increase with increasing $F^D_{\perp}$ again.  This cusp
is correlated with a sharp increase in
$P_{6}$ to $P_6=0.93$, which indicates that the system has dynamically
reordered into a triangular lattice containing a small number of fivefold and
sevenfold-coordinated defects.  We call this a 
moving hexatic (MH) state, and it exhibits
smeared sixfold peaks in $S({\bf k})$, as shown in
Fig.~\ref{fig:9}(c) at $F^{D}_{\perp} = 1.0$.
Near $F^{D}_{\perp} = 1.16$, there is another
jump in $P_{6}$ to $P_6 \approx 1.0$.  Here the system
forms a moving crystal (MC) phase,
and the corresponding structure factor in Fig.~\ref{fig:9}(d) shows much more
pronounced peaks in $S({\bf k})$.
In Fig.~\ref{fig:8}(c) we plot the velocity ratio $R$ versus
$F^{D}_{\perp}$.  There is a jump to a finite value of $R$ at the onset of the ML phase,
and a cusp at the ML-MH transition.
We do not observe any particular cusps or jumps in the transport curves at
the MH-MC transition.
In Fig.~\ref{fig:10}(a) we plot $\langle V_{\perp}\rangle$ versus
$F^{D}_{\perp}$ for the system in Fig.~\ref{fig:8}
for $A_{p} = 0.25$, 0.5, 0.75, 1.0, 1.25, and $1.5$,  with a dashed line indicating the response
for $A_{p} = 0.$
Here the cusp in $\langle V_{\perp}\rangle$ at the ML-MH/MC transition can be more
clearly seen.
Additionally, the velocity noise fluctuations are substantially reduced
in the dynamically ordered MH/MC states.
In Fig.~\ref{fig:10}(b) we show $\langle V_{\perp}\rangle$ versus
$F^D_{\perp}$ for samples with $A_{p} = 1.0$ at $\alpha_{m}/\alpha_{d} = 9.995$,
7.017, 4.925, 3.0, and $2.06$.
The MS-ML transition shifts to higher values of $F^D_\perp$ with
decreasing $\alpha_m/\alpha_d$, while the cusps at higher $F^D_\perp$
indicate the ML-MH/MC transition is still present.

The onset of different dynamical phases as a function of external driving has been
observed in various overdamped systems,
including colloids and vortices moving over q1D periodic substrates, but only for
a driving force applied parallel to the substrate periodicity direction.
In those systems there is generally a disordered flow phase above depinning
\cite{7,14,54,55} with a transition to a moving ordered phase at higher
drives \cite{14,54,55}; however, negative differential conductivity does not occur.
For systems of particles moving over 2D periodic substrates, such as egg carton or
muffin tin potentials, negative differential conductivity can arise  \cite{18,20,56,57,58}.
In previous simulations of skyrmions driven over random arrays,
it was shown that there can be a transition
from a disordered phase above depinning to a moving crystal phase
at higher drive \cite{51}, while there are extensive studies of
dynamically ordered phases as a function of increasing driving force
for vortices
driven
over random pinning arrays \cite{59,60,61,62}.

\begin{figure}
\includegraphics[width=3.5in]{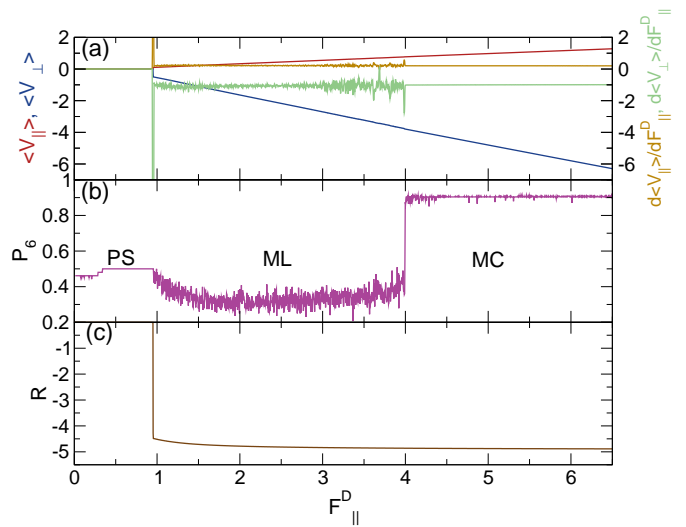}
\caption{ The same system as in Fig.~\ref{fig:8} but for driving parallel to the
  substrate periodicity direction.
(a) $\langle V_{||}\rangle$ (red) and $\langle V_{\perp}\rangle$ (blue) vs $F^{D}_{||}$,
  along with the corresponding
  $d\langle V_{||}\rangle/dF^{D}_{||}$ (yellow) and
  $d\langle V_{\perp}\rangle/dF^{D}_{||}$ (green) curves.  
  (b) $P_{6}$ vs $F^{D}_{||}$ showing transitions between the pinned smectic (PS) state,
  the moving liquid (ML), and the moving crystal (MC).
  (c)  Velocity ratio $R$ vs $F^{D}_{||}$.    
}
\label{fig:11}
\end{figure}

We have also considered the case of interacting skyrmions subjected to a drive that
is applied parallel to the substrate periodicity.
In Fig.~\ref{fig:11}(a) we plot $\langle V_{||}\rangle$ and
$\langle V_{\perp}\rangle$ versus $F^{D}_{||}$ for
a system with $\alpha_{m}/\alpha_{d} = 4.925$, $A_{p} = 1.0$, and $n_{s} = 0.16$.
Here the depinning threshold $F^{c}_{||} \approx A_{p}$, and in general for
fixed $n_s$, 
$F^{c}_{||}$ is independent of $\alpha_{m}/\alpha_{d}$
and increases linearly with $A_{p}$, similar
to the results for the single skyrmion case 
shown in Fig.~\ref{fig:2}.
The interacting skyrmions
form an immobile  pinned smectic (PS) phase which depins
plastically into a moving liquid (ML) state.
The ML dynamically orders into a moving crystal
(MC) phase near $F^{D}_{||} = 4.0$, as is illustrated by the plot
of $P_{6}$ versus $F^{D}_{||}$ in Fig.~\ref{fig:11}(b).
There is only a small cusp in the transport curves at
the ML-MC transition, as indicated by the 
$d\langle V_{||}\rangle/dF^{D}_{||}$ and $d \langle V_{\perp}\rangle/dF^{D}_{\perp}$
plots in Fig.~\ref{fig:11}(a).
This contrasts with the significantly larger cusps
that appear for perpendicular driving.
Additionally, the velocity fluctuations are strongly suppressed
once the system enters the MC phase.
The MC phase that forms for parallel driving generally contains more dislocations than the
corresponding MC phase that appears for perpendicular driving, so the parallel
driving MC phase can better be described as a moving hexatic.
In Fig.~\ref{fig:11}(c), the plot of the velocity ratio $R$ versus
$F^{D}_{||}$ shows that the PS-ML transition is sharp.
There is little
curvature in $R$ for higher drives, in contrast to the
perpendicular driving case where $R$ increases much more smoothly
as a function of drive. 

\subsection{Dynamic Phase Diagrams}

\begin{figure}
\includegraphics[width=3.5in]{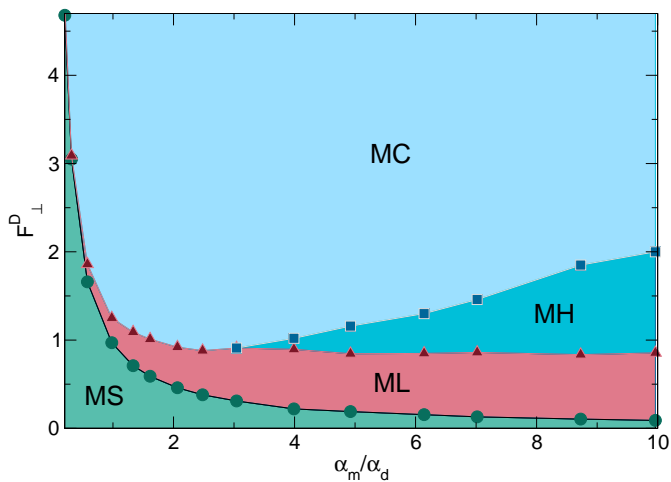}
\caption{The dynamic phase diagram as a function of $F^{D}_{\perp}$ vs
  $\alpha_{m}/\alpha_{d}$ for $A_{p} = 1.0$ and $n_{s} = 1.0$ showing
  the moving smectic (MS), moving liquid (ML), moving hexatic (MH), and
  moving crystal (MC) phases.
  Here the width of the MS phase diverges with decreasing $\alpha_{m}/\alpha_{d}$.  
}
\label{fig:12}
\end{figure}

\begin{figure}
\includegraphics[width=3.5in]{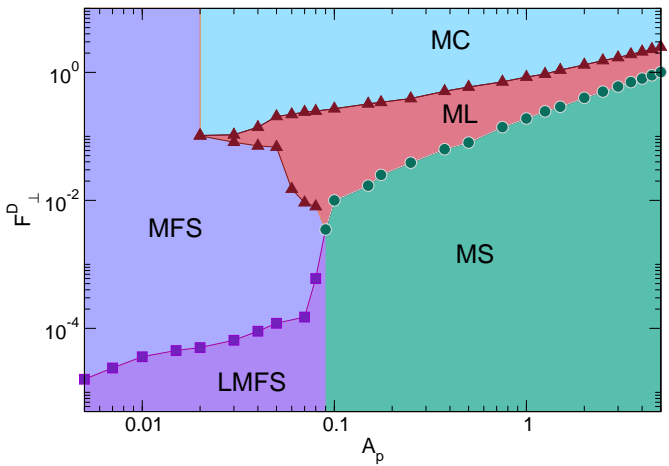}
\caption{The dynamic phase diagram as a function of
  $F^{D}_{\perp}$ vs $A_{p}$ at $\alpha_{m}/\alpha_{d} = 4.925$ and $n_{s} = 1.0$
  showing the moving solid (MS), moving liquid (ML), and moving crystal (MC) phases.
  For weaker substrate strengths, a longitudinally locked moving floating 
  solid (LMFS) phase appears which transitions with increasing
  $F^D_{\perp}$ into a phase called the moving floating solid (MFS) that moves in
  both the parallel and perpendicular directions.
}
\label{fig:13}
\end{figure}

By conducting a series of simulations and
examining the features in $\langle V_{\perp}\rangle$ and $P_{6}$, we can map
out the dynamic phases,
as shown in Fig.~\ref{fig:12}
as a function of $F^{D}_{\perp}$ versus $\alpha_{m}/\alpha_{d}$ for a system
with $A_{p} = 1.0$.
The extent of the MS phase diverges at small $\alpha_{m}/\alpha_{d}$, while the
extent of the ML phase decreases with decreasing $\alpha_m/\alpha_d$.
Based on features in the $P_{6}$ curves,
we find that the MH phase appears
only for $\alpha_{m}/\alpha_{d} > 3.0$, and that it grows in extent
with increasing $\alpha_m/\alpha_d$.
We have also considered the case of varied $A_{p}$, and in
Fig.~\ref{fig:13} we plot the dynamic phase diagram
for $F^{D}_{\perp}$ versus $A_{p}$
for a system with $\alpha_{m}/\alpha_{d} = 4.925$.
Here, the extent of the MS phase increases with increasing $A_{p}$,
and the onset of the MC phase shifts to higher $F^D_{\perp}$.
There
is also a thin strip of MH phase that appears for $A_{p} > 0.1$  (not shown).
For weak enough $A_{p}$, additional dynamical phases appear when
the skyrmions do not remain
confined to the potential minima but form  a completely triangular lattice that
is weakly coupled to the substrate.
At low $F^D_{\perp}$, the system forms a MS phase for $A_{p} \geq 0.1$,
while for $A_{p} < 0.1$ we observe a moving longitudinally locked floating solid (LMFS)
which travels strictly along the perpendicular direction.
Here, the skyrmions are pinned by the substrate in the parallel direction but can
move freely along the perpendicular direction.
At higher drives the LMFS transitions
to a moving floating solid (MFS) phase in which
the skyrmions depin from the weak substrate and begin to move
in both the parallel and perpendicular directions.

Figure~\ref{fig:13} shows that
the value of $F^{D}_{\perp}$ at which the system depins in the parallel
direction and ceases to have its motion locked along the perpendicular direction
drops substantially from the MS to the LMFS phase.
This is similar to what is observed at an Aubry transition
which arises in a 1D incommensurate Frenkel-Kontorova system, where when the substrate
is weak enough, the pinning effectively vanishes and the particles
float over the substrate \cite{63}.
It has been argued that an Aubry-like transition should occur for 2D systems
such as sliding colloids \cite{64}, and that this transition could be relevant to
the phenomenon called
superlubricity \cite{22}.
In recent simulations of colloids on 2D substrates,
it was shown that the 2D Aubry transition is
first order and is associated with a sharp drop in the effective friction \cite{64}.
This is similar to what we observe, where
there is a  sharp drop in the parallel depinning force at the MS-LMFS transition,
suggesting that this transition is first order.

\begin{figure}
\includegraphics[width=3.5in]{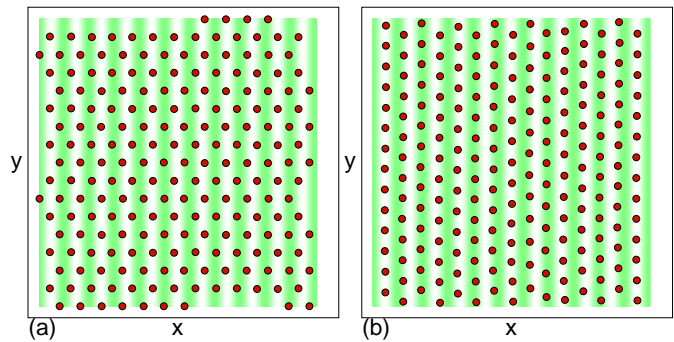}
\caption{ (a)  Image of skyrmion locations (red dots) on the substrate potential (green)
  for the system in Fig.~\ref{fig:13}
  in the moving floating solid (MFS)
  state
  at $A_{p} = 0.04$ and $F^{D}_{\perp} = 0.0001$.
  (b) The same for the moving crystal (MC) at $F^{D}_{\perp} = 1.0$,
  showing the change in the orientation of the lattice. 
 The LMFS state has the same orientation as the MFS state.
}
\label{fig:14}
\end{figure}

\begin{figure}
\includegraphics[width=3.5in]{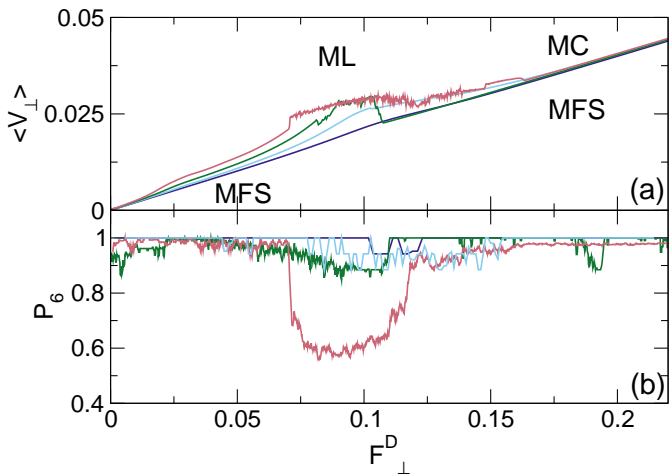}
\caption{ $\langle V_{\perp}\rangle$ vs $F^{D}_{\perp}$
  for the system in Fig.~\ref{fig:13} at $A_{p} = 0.01$ (dark blue),
  0.02 (light blue), 0.03 (green), and $0.04$ (pink). (b)
  The corresponding $P_{6}$ vs $F^{D}_{\perp}$
  showing that at $A_{p} = 0.01$ the system remains ordered,
  while at $A_{p} = 0.04$, the system goes from an ordered state into a
  disordered state and then dynamically orders at higher drives.    
}
\label{fig:15}
\end{figure}

There is a window of $A_p$ in the phase diagram of Fig.~\ref{fig:13} in which
the MFS can transition to a ML phase, which then dynamically orders into an MC
phase at higher drives, while
for low enough $A_{p}$, the ML phase is lost and no structural changes occur in the
moving skyrmion lattice as a function of $F^D_{\perp}$.
The orientation of the skyrmion lattice with respect to the substrate periodicity
direction is generally different in the MC and the MFS phases.
In the MC, the lattice is aligned with the substrate minima, while in the MFS there is
no particular matching between the lattice orientation and the substrate
periodicity direction,
as shown in the images in
Fig.~\ref{fig:14}(a,b)  at
$A_{p} = 0.04$
for $F^{D}_{\perp} = 0.0001$ and $F^D_{\perp}=1.0$.
In Fig.~\ref{fig:15}(a) we
plot  $\langle V_{\perp}\rangle$ versus
$F^{D}_{\perp}$ curves from the phase diagram in
Fig.~\ref{fig:13} for
$A_{p} = 0.01$, 0.02, 0.03, and $0.04$.
At $A_{p} = 0.01$, $\langle V_{\perp}\rangle$ is smooth,
and no change occurs in the structure of the moving triangular lattice,
while for $A_{p} = 0.02$ ,there is the beginning of a cusp feature
at $F^{D}_{\perp} = 0.1$.
For $A_{p} = 0.03$, a larger cusp appears that is associated with the
system entering the ML phase,
and a sharp drop in $\langle V_{\perp}\rangle$ occurs
when the system transitions to the moving crystal phase.
The extent of the ML phase increases for higher values of $A_{p}$,
as shown by the curve for $A_{p} = 0.04$.
In Fig.~\ref{fig:15}(b), the corresponding
$P_{6}$ versus $F^{D}_{\perp}$ curves
indicate that at $A_{p} = 0.01$,
the skyrmion lattice remains triangular with $P_{6} = 1.0$
over nearly the entire range of $F^{D}_{\perp}$.  In contrast,
for $A_{p} = 0.04$ the system transitions from
a low drive moving ordered state with $P_{6} \approx 1.0$
into the ML liquid state, as shown by the drop to
$P_{6} \approx 0.6$.  This is followed at higher drives by
a transition into the moving crystal phase where
$P_{6} \approx 1.0$ again.

\section{Varied Skyrmion Densities}

\begin{figure}
\includegraphics[width=3.5in]{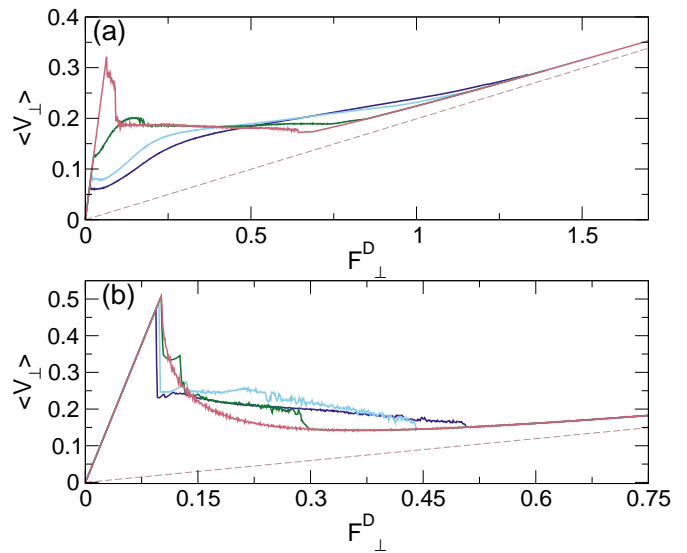}
\caption{$ \langle V_{\perp}\rangle$ vs $F^{D}_{\perp}$
  for samples with $\alpha_{m}/\alpha_{d} = 4.925$
  and $A_{p} = 0.5$ at varied skyrmion densities  $n_s$.
  (a) $n_{s} = 0.44$ (dark blue), 0.36 (light blue), 0.262 (green), and $0.208$ (pink).
  The dashed line indicates the response for $A_{p} = 0.$
  (b) $n_{s} = 0.00926$ (pink), 0.023 (green), 0.0612 (light blue), and $0.129$
  (dark blue).
  The dashed line indicates the response for $A_{p} = 0.$
}
\label{fig:16}
\end{figure}

\begin{figure}
  \includegraphics[width=3.5in]{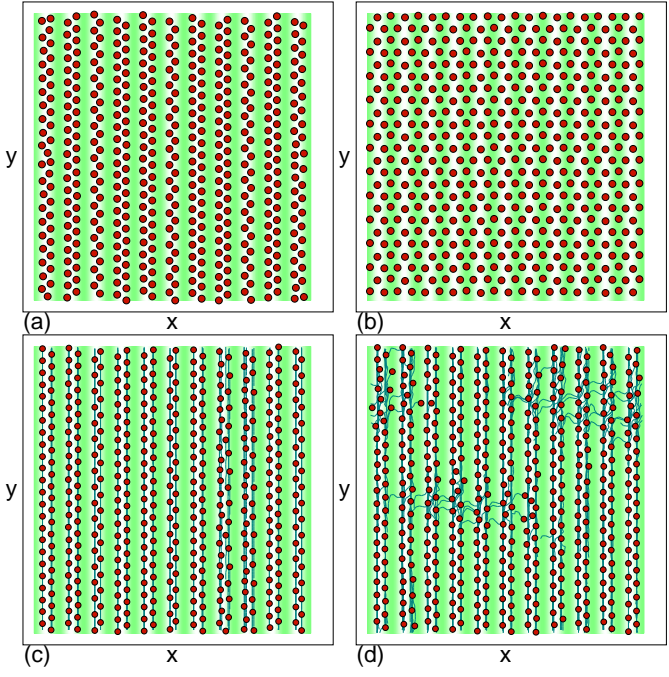}
  \caption{
  Images of skyrmion locations (red dots) on the substrate potential (green) for the system
  in Fig.~\ref{fig:16} with $A_p=0.5$ at $n_s=0.37$.
    (a) Snapshot of the pinned state at $F^D_{\perp}=0$ where there are two
    rows of skyrmions per potential minima.
    (b) Snapshot of the moving crystal state at $F^D_{\perp}=0.5$.
    (c) Skyrmion trajectories in the MS phase with two rows of moving skyrmions per
    potential minima at $F^{D}_{\perp}=0.02$.
    (b) Skyrmion trajectories showing
    coexisting MS and ML flow at $F^{D}_{\perp} = 0.08$. 
}
\label{fig:17}
\end{figure}

We next consider the effect of varying the skyrmion density for
perpendicular driving.
We expect that a series of commensurate and incommensurate transitions should
occur as a function of the ratio of the skyrmion lattice spacing to the substrate lattice
constant, as observed for superconducting vortex and colloidal systems; however, a
study of such effects is outside the scope of the present work.
In Fig.~\ref{fig:16}(a) we plot $\langle V_{\perp}\rangle$ versus
$F^{D}_{\perp}$
for a system with $\alpha_{m}/\alpha_{d} = 4.925$
and $A_{p} = 0.5$ at $n_{s} = 0.44$, 0.36, 0.262, and $0.208$.
The dashed line is the result for $A_{p} = 0$, which is independent of $n_s$.
The $n_{s} = 0.208$ results are very similar to
the behavior at $n_{s} = 0.16$ shown in Fig.~\ref{fig:8}, where
there is  drop in $\langle V_{\perp}\rangle$ at the MS-ML transition
and a cusp at the ML-MC transition.
As $n_{s}$ increases,
the extent of the MS phase decreases
and the onset of the MC phase shifts to higher
values of $F^{D}_{\perp}$.
For the higher values of $n_{s}$, the MS phase contains multiple rows of moving
skyrmions per substrate minimum, as illustrated
in Fig.~\ref{fig:17}(a) at $n_{s} = 0.37$ and $F^{D}_{\perp} = 0$.
Fig.~\ref{fig:17}(c) shows the particle trajectories in the moving locked phase at
$F^D_{\perp}=0.02$ where the motion occurs in one-dimensional channels.
The $\langle V_{\perp}\rangle$ curves in Fig.~\ref{fig:16} also show that the speed up effect
observed in the single skyrmion limit remains
robust when the skyrmion density increases.
Just above the MS-ML phase transition for
the $n_{s} = 0.44$, 0.37, and $0.262$ curves in Fig.~\ref{fig:16},
there is a region in which a coexistence of MS and ML flow occurs,
as illustrated in Fig.~\ref{fig:17}(d) for 
$n_{s} = 0.37$ and  $F^{D}_{\perp} = 0.08$.
At high drives the system can dynamically order into the moving crystal phase, as
shown in Fig.~\ref{fig:17}(b) for $n_s=0.37$ and $F^D_{\perp}=0.5$.
In Fig.~\ref{fig:16}(b) we plot $\langle V_{\perp}\rangle$
versus $F^{D}_{\perp}$ for the same system at
$n_{s} = 0.00926$, 0.023, 0.0612, and $0.129$,
where the dashed line shows the
curve for a sample with $A_{p} = 0.$
As $n_{s}$ decreases, the extent of the
ML phase is reduced
while the ML-MS transition point remains almost constant.
For $n_{s} = 0.00926$, the $\langle V_\perp\rangle$ curve is almost the same
as that found for the single skyrmion case,
and there is no clear ML-MC transition.

\begin{figure}
\includegraphics[width=3.5in]{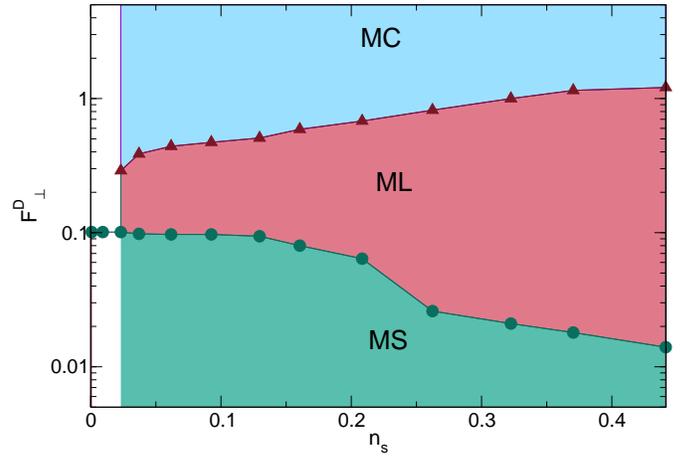}
\caption{ The dynamic phase diagram for the system in Fig.~\ref{fig:16}
  as a function of $F^{D}_{\perp}$ and $n_{s}$. 
  The MS-ML transition drops to lower values of $F^D_{\perp}$ with
  increasing $n_{s}$ once double rows of skyrmions can form in the
  potential minima, as illustrated
  in Fig.~\ref{fig:17}(a).
  For $n_{s} < 0.02$, the system behavior is the same as in the single skyrmion limit.  
}
\label{fig:18}
\end{figure}

In Fig.~\ref{fig:18} we map the dynamic phase diagram
as a function of $F^{D}_{\perp}$  and $n_{s}$.  The
MS-ML transition line drops markedly above
$n_{s}= 0.25$ due to the formation of 
double rows of skyrmions in each substrate minimum in
the MS phase,  as illustrated in Fig.~\ref{fig:17}(a).
Additionally, for $n_{s} < 0.02$ the
system behavior becomes identical to the single skyrmion limit.
These results show that the skyrmion phases we observe should be robust
over a wide range of magnetic fields.

\begin{figure}
\includegraphics[width=3.5in]{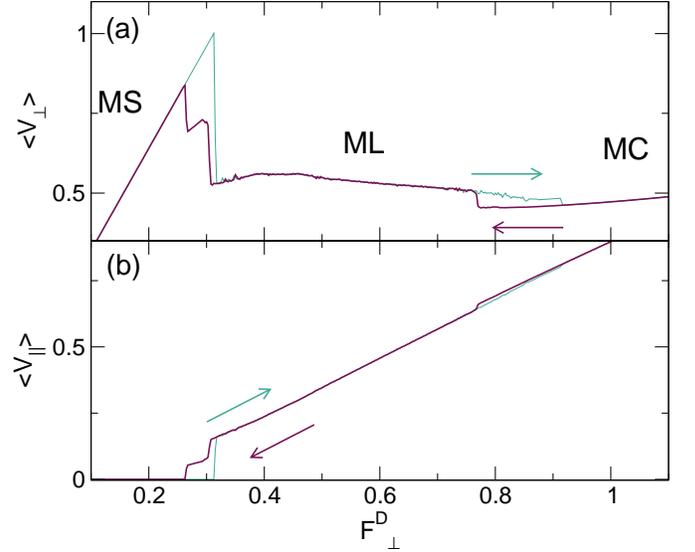}
\caption{ 
  (a) $\langle V_{\perp}\rangle$ vs $F^{D}_{\perp}$ for
  a system with $n_{s} = 0.16$, $A_{p} = 1.0$, and $\alpha_{m}/\alpha_{d} = 3.042$.
  The thin green line indicates the ramp up and the thick red line indicates the ramp down,
  showing hysteresis across the MH-ML and ML-MS transitions.
 (b) The same for $\langle V_{||}\rangle$ vs $F^{D}_{\perp}$.
}
\label{fig:19}
\end{figure}

We have also examined hysteretic effects across the different dynamic phase
transitions
by ramping the applied drive up and down, as shown in Fig.~\ref{fig:19} where
we plot 
$\langle V_{\perp}\rangle$ and $\langle V_{||}\rangle$
versus $F^{D}_{\perp}$ at $A_p = 1.0$, $n_{s} = 0.16$, and
$\alpha_{m}/\alpha_{d} = 3.042$.
The thin line is the ramp up curve and the thick line is the ramp down
curve.  Hysteresis appears 
in both $\langle V_{||}\rangle$ and $\langle V_{\perp}\rangle$
across the ML-MC phase transition, where the system remains
locked in the moving crystal phase down to lower drives than that at which the MC
phase first appears on the increasing portion of the ramp.
There is also hysteresis across the ML-MS phase transition, where the system
remains in the ML phase down to lower drives for the ramp down than
during the ramp up.
In general, we observe hysteresis for all values of $\alpha_{m}/\alpha_{d}$, with
a slight increase in the width of the
hysteretic intervals for increasing $\alpha_{m}/\alpha_{d}$.
This shows that several of the dynamic phases have first order like features,
and that hysteresis in the transport curves provides
another method for identifying the onset of the different dynamic phases.

\section{Summary}
We have examined individual and multiple skyrmions
in a 2D system driven over a quasi-1D periodic substrate where the Magnus term in the
dynamics produces new effects that are not observed in overdamped realizations
of this geometry.
When the driving force is applied parallel to the substrate
periodicity direction, the depinning force is not reduced when the magnitude of
the Magnus force increases,
in contrast to what occurs for pointlike pinning.
This is because the planar nature of the quasi-1D pinning substrate does not allow the
skyrmions to curve around and avoid the pinning sites, as is possible for pointlike pinning.
For driving in the direction perpendicular to the substrate periodicity, 
in the overdamped limit the substrate potential has no effect
and the velocity force curves are linear as the particles simply slide
along the substrate.   When a finite Magnus term is present, however,
a rich variety of dynamical
effects can arise.  At lower external drives the skyrmion
motion is locked to the direction of drive,
while at higher drives there is a transition to motion both transverse and
parallel to the applied drive.
At this transition there is a decrease in the net skyrmion velocity,
producing a negative differential conductivity effect.
Within the longitudinally locked phase, there is a pronounced speed up effect
in which the skyrmions move faster than particles in the overdamped limit would move.
This occurs when the
velocity component from the Magnus term
is aligned with the external drive. Such speed up effects were previously observed for
systems with pointlike or circular pinning sites.
Here we find a speed up effect only for perpendicular driving.
For collectively interacting skyrmions, a variety of distinct dynamical phases arise
including moving smectic, liquid, hexatic, and crystal phases.
The transitions into and out of many of these phases
produce dips and cusps in the transport curves.
We map the onset of these different phases as a function of the ratio
of the Magnus term to the dissipative term, the external drive, the substrate strength, and
the skyrmion density. For varied substrate strengths we find
evidence for an Aubry like transition when the substrate is
weak enough that the skyrmions form a floating triangular solid.
For increasing skyrmion density, we find a transition from one to multiple rows of skyrmions
in each substrate minima, which coincides with a decrease in the
range of driving force values over which skyrmion
motion remains locked in the direction of the perpendicular driving force.
A potential of the type we consider
could be realized using samples with periodic thickness modulations
or magnetic line pinning, or even via optical means,
and  the existence of different skyrmion dynamical phases
could be deduced from changes in the transport curves or
by observing dynamical changes of the skyrmion configurations.

\acknowledgments
This work was carried out under the auspices of the 
NNSA of the 
U.S. DoE
at 
LANL
under Contract No.
DE-AC52-06NA25396.

\end{document}